\documentclass[11pt]{article}%
\usepackage{amssymb}
\usepackage{amsfonts}
\usepackage{amsmath}
\usepackage[nohead]{geometry}
\usepackage[singlespacing]{setspace}
\usepackage[bottom]{footmisc}
\usepackage{indentfirst}
\usepackage{endnotes}
\usepackage{graphicx}%
\usepackage{rotating}
\usepackage{color}
\newtheorem{theorem}{Theorem}

\newtheorem{hypothesis}{Hypothesis}

\newtheorem{definition}[theorem]{Definition}

\newtheorem{assumption}{Assumption}

\newtheorem{proposition}{Proposition}

\newenvironment{proof}[1][Proof]{\noindent\textbf{#1.} }{\ \rule{0.5em}{0.5em}}

\makeatletter
\def\@biblabel#1{\hspace*{-\labelsep}}

\makeatother
\begin{document}

\title{Coevolution of a network and perception\thanks{We are grateful to Robin Dunbar, Martin Everett, and Stephan Pfaffenzeller for valuable comments.}
}
\author{Hang-Hyun Jo\thanks{Department of Physics, Pohang University of Science and Technology, Pohang 790-784, Republic of Korea and BECS, Aalto University School of Science, P.O. Box 12200, Finland, johanghyun@postech.ac.kr} \quad Eunyoung Moon\thanks{Corresponding author. Address: University of Liverpool Management School, Chatham Street, Liverpool L69 7ZH, UK. Phone: +44 0151 795 3530. Email: E.Moon@liverpool.ac.uk  }}
\maketitle

\sloppy%avoids the breakage of words at the end of lines, by adjusting spaces between words inside the lines

\onehalfspacing

\begin{abstract}
How does an individual's cognition change a system which is a collective behavior of individuals? Or, how does a system affect an individual's cognition? To examine the interplay between a system and individuals, we study a cognition-based network formation. When a network is not fully observable, individuals' perception of a network plays an important role in decision making. Assuming that a communication link is costly, and more accurate perception yields higher network utility, an agent decides whether to form a link in order to get better information or not. Changes in a network with newly added links affect individuals' perception accuracy, which may cause further changes in a network. We characterize the early stage of network dynamics and information dispersion. Network structures in a steady state are also examined. Additionally, we discuss local interactions and a link concentration in a frequently changing network.
\end{abstract}

\strut


\strut


\pagebreak
\doublespacing

In the 1960's, Stanley Milgram conducted a notable experiment to examine the average path length in a social network. In his experiment, a randomly selected person was asked to pass a packet to either a target or someone in his/her acquaintance who might be most likely to know the target. The result of this experiment, widely known as \emph{six-degrees of separation}, shows how close people are in a social network. Since his experiment, studies of distance in a network have had an important and long-standing research tradition in network theory. From empirical studies (Sampson, 1998; Hedstr\"{o}m, Sandell, and Stern, 2000; Newman, 2001; Kretschmer, 2004; Kossinets and Watts, 2006) to theoretic approaches (Watts and Strogatz, 1998; Watts, 1999), a wide range of research has examined distance in a network. However, the research has not paid attention to another aspect of this experiment - cognition of a social network: If participant A forwarded the packet to his/her friend B, why did he/she choose B, why not another friend C? Following the experiment instruction, it is simply because A thought that B is closer to the target than C. In other words, in A's perceived network, B is the node which has the shortest path length to the target.

In this experiment, the information used by agents for decision making is not the real network but an individual's own cognition of the real network. Generally, when people do not observe the whole network structure, perception of reality is more influential in individuals' decision making than the reality itself. Nonetheless, to the best of our knowledge, there is no theory of network formation which takes account of perception in decision making, and this motivates us to suggest a cognition-based network formation model.

\smallskip

In this research, we focus on an ecology of how a social network as a system of collective behaviors evolves with individuals' cognition, questioning how people make a connection, how individuals' perception converges into reality, and what the stable network structure will be. Precisely, we consider a cognition-based strategic network formation model: In a reasonably large group,\footnote{As Hill and Dunbar (2003) empirically showed that the human brain can cope with a limited number of social relations (approximately maximum 150 on average), unlike other growing network models (Barab\'asi and Albert, 1999; V\'{a}zquez, 2003), we fix the number of nodes and examine how these nodes create links to each other. In the simulation, we set 200 nodes which is natural to assume that everyone knows the existence of others in the group regardless of communication links.} people may not fully observe the entire relationships within a social network, instead, they have a perception of the network. To describe each individual's perceived network, we follow the concept of ``the cognitive social structure (CSS)'' defined by Krackhardt (1987)\footnote{He attempted to aggregate individuals' cognition in order to derive a representative CSS. In our model, we keep individual agents' cognition in $n\times n\times n$ matrices for $n$ nodes.} as the relation between a sender, a receiver, and a perceiver. We also introduce the notion of perception accuracy, which captures the number of correctly perceived links in all possible relationships. Without full information of the network structure, how well one can utilize a network depends on the accuracy of one's perception.\footnote{Network utility in this model is distinguishable from other strategic network formation models in which utility comes from a link itself. The utility here is oriented not from links but from the information about links, capturing that people use a social network as much as they know it.} Given a network utility as an increasing function of the perception accuracy, people infer others' accuracy by observing their utility, thus we can consider the network utility as a proxy measure of the perception accuracy. It also leads to individuals' incentive for linking to more accurately perceiving agents in order to improve their perception. Since linking is costly, a link will be created only if the advantage of having more accurate perception exceeds the cost of linking. On the other hand, whenever a link is added to a current network, the newly added link affects individuals' accuracy, hence both the advantage of high accuracy and the disadvantage of low accuracy are temporary as long as a network continues evolving. Starting with an empty network in reality and an Erd\H{o}s-R\'{e}nyi random network with probability $p$ in cognition, the network dynamics ends when no one would like to add a link.

\smallskip

In Section 1, we briefly summarize related discussions and our results. Section \ref{model} presents the model and analysis, and Section \ref{simulation} illustrates results in Section \ref{model} by an agent-based simulation. Concluding remarks are offered in Section \ref{conclusion}.
\bigskip

\section{Background and overview of results}

\textbf{Network formation.} There are plenty of studies on network formation models. First of all, stochastic network formation models have provided understanding of important aspects of evolving structure, starting from the seminal papers about mechanisms behind the small-world phenomena. In particular, Barab\'asi and Albert (1999) showed that a hub node emerges in a growing network by preferential attachment process. In their model, since the probability of getting a new link is increasing in the number of existing links, the one who is more connected has the higher chance to get linked so that a network ends up with a highly centralized structure such as a scale-free network.

On the other hand, a game theoretic approach emphasizes the importance of decision making in network formation. Since linking is costly, individual agents optimize their connections, thus a network structure depends on the level of cost. In particular, two extreme structures arise such as the empty network under the high cost and the complete network under the low cost, and a star network appears at the moderate level of cost  (Jackson and Wolinsky, 1996; Bala and Goyal, 2000; Watts, 2001).

Note that for either stochastic network models or strategic network formation, an agent knows the whole structure of a network when a link is formed. Unlike previous studies, our model is interested in the case where full information about a network is unavailable: we add a cognitive perspective to the current network literature by developing a cognition-based strategic network formation model.

\bigskip

\textbf{Cognition in a network.} While cognitive aspects in network formation have not been discussed, discussion about the relationship between size/layers of networks and human brain capacity has been relatively active (Rose and Serafica, 1986; Dunbar, 1998; 2009; Stiller and Dunbar, 2007; Roberts and Dunbar, 2011; Sutcliffe, Dunbar, Binder, and Arrow, 2012). Known as ``social brain hypothesis'', there is biological evidence to show a relationship between the size of social groups and brain. Regarding our question of cognition-based network formation, those works can support the cost of linking in strategic link formation if the capacity of a social brain related to the size and depth of a social network is interpreted as a cost. That is, the brain capacity of socializing can be a specific type of cost of linking.

There are a few works more directly related to this research: since Krackhardt (1987) has formalized individuals' perception of a network firstly, Krackhardt and Kilduff (1999; 2008) have discussed how people perceive social networks and how network cognition affects an individual's behavior in a network.  Especially, recent experimental approaches to a cognition bias reveal that people perceive closer and denser than reality (Kilduff, Crossland, Tsai, and Krackhardt, 2008) and  perceived network structures are flatter than reality (Dessi, Gallo, and Goyal, 2012).  Further than the disparity between cognition and the real network, we examine how an individual's perception converges to reality and how cognition and a network coevolve.

\bigskip

\textbf{Local interaction.}  Although triadic relations are widely observed in a social network, there are few theories about the formation of triadic relations. Firstly, Granovetter (1973) explains that two strongly connected agents are more likely to form an interaction circle with another agent. On the other hand, a recursive search model (V\'{a}zquez, 2003) and network-based meetings (Jackson and Rogers, 2007) show the influence of existing links on the new link formation by combining random connections with local search. That is, since one connects with another through existing connections, any two agents who share a common neighbor are more likely to be connected, and this tendency is stronger if the already existing tie is strong. These studies emphasize a mediating node which plays a bridging role to form a cluster.

Our model provides a different explanation for clustering which does not need a mediator: the closer the distance between %the most/least
the most and the least accurate agents in a network, the more triadic relations appear.

\bigskip

\textbf{Main results and contribution.} As mentioned, most network formation models tend to take full information about a network structure for granted, and we try to expand the discussion into how people perceive a network. In this research, people use their perceived network rather than the real network for link formation, and their cognition-based networking changes the real network structure, leading to changes in perception back. Our model provides a theoretic framework for coevolution of a network and cognition: we derive the early stage of an evolving network and behavioral hypothesis and illustrate the model by an agent-based simulation.  We briefly introduce key findings as follows.

Firstly, a network structure in the early stage of evolution is a ring, which implies that a dominantly centralized agent may not exist. In the further evolving process, a network evolves either in a complex structure by short cuts or in a global ring structure with local connections.

\smallskip

Secondly, since everyone knows his/her own link states correctly, the correct information is added on the updated perception until full information is completed. The full information with a small fluctuation caused by newly added links spreads, thus the gap between the most and the least accurate perception decreases, which weakens the incentive for a new link.

\smallskip

Thirdly, the cost of linking and the network evolution are negatively related. We find discontinuous jumps in terms of stable link density in accordance with the cost of linking: high cost leads to an empty network in a steady state, low cost results in a complex network, and a ring network arises at the moderate level of cost.

\smallskip

Additionally, we discuss how the frequency of linking affects a network structure. In particular, a clustering mechanism without a mediating node and a link concentration are investigated under a multiple agents' perception update circumstance. If several agents update their perception in one time period, the most and the least accurate agents are closely located in a network. New links are added between closely located agents and a triadic relation arises in local connections. Moreover, an agent who is slightly more connected than others by chance grows into a hub due to the exclusive information about newly added links.

\smallskip

From these findings, our paper contributes to network theory in three ways. Firstly, we believe that this is the first work to explain network formation based on an individual's perception. This model is established on ``what people are really aware of'' rather than on ``what reality is''. Secondly, our findings on network formation improve our understanding of the interplay between individuals and a system. The result shows the complementarity between perception and a network structure such that differences between perception and reality trigger changes in a network and the structural changes accelerate the gap which leads to more changes in a network. Finally, since this model suggests a general framework, there is potential applicability to various fields. For instance, any communication related networks such as coevolution of rumors and friendship in sociology and corporate governance and directorship in finance can be examined by this model.

\bigskip

\section{The model}
\label{model}

We consider a two-way flow communication network and an individual's perception of the network.  Communication for information transmission occurs between directly connected agents, and each individual has his/her own perception of who connects with whom. A network in one's perception does not necessarily coincide with the actual network, and it is natural to assume that the one who perceives a network more accurately can use it better. Although perception details are not observed unless directly shared via a communication link, how well an agent utilizes a network reveals his/her overall accuracy of the perception. Thus, perception itself is private information which can be shared via a link, whereas the accuracy of perception is observable to all agents by network utility.

Once individual agents observe others' accuracy, less accurate agents may want to know the most accurate perception details in order to improve the network utility. If an agent has a link to the most accurate agent, he/she uses the existing link for the perception update. If an agent does not have a link to the most accurate one, he/she needs to create a new link which is costly. Hence, the least accurate agent is willing to link to the most accurate agent only if the additional utility by updating perception exceeds the cost of linking.

We are interested in how the network structure and agents' perception evolve together. We firstly analyze a network in the early stage and an individual's perception update process, then characterize an evolving network structure.

\bigskip

\subsection{Settings}

\textbf{Network.} Let $N=\{1, \cdots, n\}$ be the set of individuals. In order to exclude perception updates without communication,\footnote{For instance, consider three agents in a line network. Each agent correctly knows his/her link states with the other two and there is only one uncertain link state between the other two, thus it is possible to reach the full information without communication.} we assume that $n$ is a sufficiently large number. At the same time, when it comes to the cognitive problem, $n$ cannot be a huge number in order for everyone in a network to recognize the existence of $n-1$ others regardless of the link states. We open $n$ as an arbitrary number.

For any pair of $j, k \in N$, $e_{jk, t}$ represents the relation between $j$ and $k$ in time period $t$.  When $j$ and $k$ are connected, $e_{jk,t}=1$, while $e_{jk, t}=0$ refers to the case of no connection. For a two-way communication link, $e_{ij}=e_{ji}$. By convention, $e_{ii}=0$ for all $i\in N$. A network $G_t$ is a collection of link states at $t$, i.e., $G_t=\{e_{jk,t}\}_{j,k\in N}, t=0, 1, \cdots, T$. Degree of $i$, denoted by $d^i_t$, is the number of $i$'s links in $G_t$, i.e., $d^i_t\equiv\sum_{j\in N}e_{ij,t}$.

We consider a time-evolving network $G_t$ in which only one link can be added in each time period.\footnote{In an evolving network, to whom to allow to form a link results in different network structures. For the simplest case, if there is no restriction on adding a link, all links are concentrated to the most favorable agent so that a star network arises. On the other hand, if a new link chance is prioritized, non-trivial structures may arise. Once we introduce all necessary notations and settings, we will discuss about it more deeply.} Denoting $G+ij$ as a network $G$ with a new link between $i$ and $j$,
\begin{equation}
    G_{t+1}= \begin{cases} G_t+ij &\mbox{if } e_{ij,t+1}=1 \text{ for }i, j \text{ such that }e_{ij,t}=0 \\
G_t & \mbox{otherwise. } \end{cases} \nonumber
\end{equation}

Individuals do not observe the actual network $G_t$ except their own link states to $n-1$ others, instead, each of them has their own perception of $G_t$.

\bigskip

\textbf{Perception, accuracy and network utility.} In the same way of defining a network $G_t$, individual agents' perception on the network $G_t$ can be defined as $G^i_t=\{e^i_{jk, t}\}$ for $i \in N$. The network $G^i_t$ refers how $i$ perceives the actual network: if $i$ thinks that $j$ and $k$ are connected, $e^i_{jk,t}=1$, while $e^i_{jk, t}=0$ implies that $i$ thinks there is no link between $j$ and $k$. The perception $G^i_t$ is private information which can be shared via a link.

Now we measure how accurate one's perception is.
\begin{definition} Perception accuracy of $i$ %
  is defined as
\begin{equation}
    \rho^i_t= \frac{1}{M}\sum_{j \in N }\sum_{k \in N, k >j}I(e_{jk,t}, e^i_{jk,t}), \quad \frac{2}{n} \leq \rho^i_t \leq 1,\nonumber
\end{equation}
where $M=\frac{n(n-1)}{2}$ denotes the number of all possible pairs among $n$ agents, and $I(x,y)$ is an index function, having a value of $1$ if $x=y$, otherwise $0$.
\end{definition}
The accuracy $\rho^i_t$, which is the aggregated information of $M$ link states in $G^i_t$, captures $i$'s correct information out of all possible pairs. Note that since $i$ correctly knows at least $n-1$ link states related to $i$ itself (i.e. $e^i_{ij}=e_{ij}$ for all $j\in N\setminus\{i\}$), the lower bound of the accuracy is $\frac{2}{n}$. Being closer to 1 implies more accurate perception.

\smallskip

In the beginning of each time period, individual agents' accuracy is adjusted to the actual network. Precisely, for $i$ who keeps the same perception in the next time period ($G^i_t=G^i_{t+1}$), its accuracy in $t+1$ is
\begin{align}
  \label{rhoUpdate}
\rho^i_{t+1}=\frac{1}{M}\sum_{j \in N }\sum_{k \in N, k >j}I(e_{jk,t+1}, e^i_{jk,t})=\begin{cases}\rho^i_t &\mbox{ if } G_{t+1}=G_t\\\rho^i_t\pm \frac{1}{M} &\mbox{ if } G_{t+1}\neq G_t.\end{cases}
\end{align}\footnote{Observe that the accuracy can be either improved or declined because of newly added links. Consider three agents $i, j, k$ such that $e_{jk,t}=0$. If $e^i_{jk,t}=0$ and $j$ creates a link to $k$ in $t+1$, $i$'s accuracy in $t+1$ is $\rho^i_{t+1}=\rho^i_{t}-\frac{1}{M}$. Oppositely, if the original perception was wrong ($e^i_{jk,t}=1$), $i$'s accuracy is improved ($\rho^i_{t+1}=\rho^i_{t}+\frac{1}{M}$) by an accidental correction.}

\smallskip

Since individuals utilize a network $G_t$ as much as they know it, we define network utility as a function of an individual's accuracy. To reflect the intuition that the more accurate information leads to the better use of a network, assume that network utility is increasing in the accuracy as follows:$$u_i(G_t|G^i_t)=u(\rho^i_t)\quad \forall i\in N,\quad u'(\cdot)>0.$$ Note that although $i$'s perception $G^i_t$ is private information, the accuracy of $G^i_t$ is indirectly observed by the network utility.\footnote{Since $\rho^i_t$ is the aggregated information of $G^i_t$, even if $\rho^i_t$ is revealed by the network utility, details in $G^i_t$ is unobserved to others unless $i$ directly shares via a communication link.} Moreover, since the actual network $G_t$ is unobserved, knowing $\rho^i_t$ by $i$'s own utility does not imply that $i$ knows which link states in $G^i_t$ are correct or wrong.

\bigskip

\textbf{Perception updates.} Since a link is a conduit of communication to make individuals share their own perception with others, the network utility can be improved if one uses a link to obtain more accurate perception.
Considering the strongest incentive for improving perception accuracy, we assume that in each time period, a chance for the perception update is given to the least accurate agent. Moreover, it is obvious that the least accurate agent would like to communicate with the most accurate agent for the best use of the update chance.
\begin{assumption}\label{assume}
In time period $t$, a chance for the perception update is given to $l_t$ such that $$l_t\in L_t\equiv \{\forall i|\arg \min_N (\rho_t^1, \cdots, \rho_t^n)\}.$$ % must be a set

If $|L_t|>1$, $l_t$ is a randomly chosen element in $L_t$.

$l_t$ updates its perception using $h_t$'s perception as follows:
\begin{equation}
  e^{l_t}_{jk,t+1}=\begin{cases} e^{h_t}_{jk,t} & \mbox{ if }j,k\in N\setminus\{l_t\}\\
    e^{l_t}_{jk,t} &\mbox{ otherwise}, \end{cases}
  \label{update}
\end{equation}
where $h_t$ is the most accurate agent such that
$$h_t\in H_t\equiv \{\forall i|\arg \max_N (\rho_t^1, \cdots, \rho_t^n)\}.$$ If $|H_t|>1$ and $\mathcal{H}_t\neq \varnothing$, where $\mathcal{H}_t\equiv \{\forall i|e_{l_t i,t}=1, i\in H_t \}$, %and $\mathcal{H}_t \subset H_t$,
$h_t$ is a randomly chosen element in $\mathcal{H}_t$. If $|H_t|>1$ and $\mathcal{H}_t= \varnothing$, $h_t$ is a randomly chosen element in $H_t$.
\end{assumption}
The least and the most accurate agents in time period $t$ are denoted by $l_t$ and $h_t$ respectively. For a simple notation, we omit the subscript $t$ of $l_t, h_t$ unless absolutely necessary.

In Assumption 1, $l$, which has the lowest accuracy in $t$, finds the most accurate agent $h$ and obtains $h$'s perception details. Since $\rho^h_t$ does not reveal whether a specific link state is correct or not, the best way of $l$'s update is replacing its own perception on all link states with $h$'s perception except $n-1$ link states related to $l$ itself, as specified in (\ref{update}).

Note that by this assumption, we confine the perception update to one agent in each time period. However, since there can be several agents who have the lowest accuracy and allowing all of them to update could show a different intuition, we relax this assumption in Extension section by allowing all least accurately perceiving agents to update their perception. Additionally, if the perception update is allowed to anyone who is willing to update, a star network arises.\footnote{Suppose that several individuals want to update their perception with the originally most accurate agent's perception $G^h_t$. If all of them update and some of them form a link to $h_t$ for the update, $h_t$ still has the highest accuracy in the next time period because the newly formed links to $h_t$ is fully known to $h_t$ only. If some agents want to update in the next time period, they still choose the most accurate agent so that $h_t$ will be selected again so that all links will be concentrated on $h_t$.} We omit this case to avoid a trivial analysis.

\smallskip

To obtain $h$'s perception details, $l$ needs a link to $h$. If $l$ already has a link to $h$ in time period $t$ (i.e. $e_{lh,t}=1$), communication for the perception update is not costly. If $e_{lh,t}=0$, $l$ needs to create a link to $h$, which is costly. We set an arbitrary non-negative cost of linking $c\geq0$ which is imposed only on the one who suggests a link ($l$), not on the one who gets a link offer ($h$).\footnote{We emphasize that the cost of linking is an effort to initiate a relation. It implies that severance of existing links does not occur because the cost of maintenance is out of the scope in this model. It is possible way to expand this model into the cost of maintenance by setting the upper bound of degree.}

In the next time period $t+1$, $l_t$'s perception accuracy is
\begin{equation}
    \rho^{l_t}_{t+1}= \begin{cases} \frac{1}{M}\sum_{j\in N}\sum_{k\in N, k>j}I(e_{jk,t+1},e^{l_t}_{jk,t})=\rho^{l_t}_t &\mbox{if no update}  \\
\frac{1}{M}\left[n-1 + \sum_{j\in N\setminus\{l_t\}}\sum_{k\in N\setminus\{l_t\}, k>j}I(e_{jk,t+1}, e^{h_t}_{jk,t}) \right]=\frac{2}{n}+\left(1-\frac{2}{n}\right)\rho^{h_t}_t & \mbox{if update.}  \end{cases} \label{rho}
\end{equation}
Observe that $l_t$ has more accurate perception than $h_t$ in the next time period $t+1$ by updating (i.e. $\frac{2}{n}+\left(1-\frac{2}{n}\right)\rho^{h_t}_t\geq \rho^{h_t}_t$).
The first case of (\ref{rho}) is that $l$ does neither have nor create a link to $h$. Since creating a new link is costly, the second case of (\ref{rho}) occurs only if the benefit of improving the perception accuracy by a new link exceeds the cost of linking. That is, a new link will be added if and only if
\begin{equation}
u\left(\frac{2}{n}+\left(1-\frac{2}{n}\right)\rho^{h_t}_t\right)-c > u(\rho^{l_t}_t) \quad \Leftrightarrow \quad u\left(\rho^{l_t}_{t+1}\right)- u(\rho^{l_t}_t)>c. \label{newlink}
\end{equation}
For a linear utility $u(\rho)=\rho$ as a simple example, the condition for a new link in (\ref{newlink}) is
\begin{equation}
    \frac{2}{n}+\left(1-\frac{2}{n}\right)\rho^{h}_t-\rho^{l}_t >c.
    \label{case_linear}
\end{equation}
From now on, we keep the linear utility $u(\rho)=\rho$ for simplicity. As long as the assumption $u'(\rho)>0$ holds, the curvature of a utility function only affects the time when a network stops evolving: Given $c>0$, if a utility function is concave, the incentive to add a link is stronger for lower $\rho^l_t$ and weaker for higher $\rho^l_t$ because the increment of utility by improving accuracy is higher at the low level of $\rho_t$. Similarly, for a convex utility function, the higher $\rho^l_t$ has a stronger incentive for adding links. In the analysis, we will firstly investigate a network structure and information spreading in the early stage, to which the shape of a utility function is irrelevant. For an analysis of a steady state, we still retain the linear utility function as a benchmark.

\bigskip

\textbf{Decision flow.} In the beginning of each time period $t$, the network $G_t$ and perception $G^i_t$ for all $i\in N$ are given. Individuals firstly observe the accuracy profile $(\rho^{1}_t, \cdots, \rho^{n}_t)$. The second step is the perception update decision of $l_t$. If necessary, $l_t$ decides whether to create a link. Once the perception update decision has been made, time period $t$ ends. In the beginning of $t+1$, all agents' perception accuracy is adjusted according to (\ref{rhoUpdate}) and then the accuracy profile $(\rho^{1}_{t+1}, \cdots, \rho^{n}_{t+1})$ is revealed. All steps are repeated until no one would like to update perception.

\bigskip

\subsection{Analysis}

In this section, we show how individuals make a decision to connect, how perception affects a link formation, and how the entire network structure evolves. Since the cost $c$ only plays a partial role for the network structure by affecting the time when a network reaches a steady state, we ignore the cost ($c=0$) in order to identify the early stage of an evolving network, and then consider network structures in a steady state together with the cost. To see a structural change, we start with an empty network in reality, while individuals perceive the real network as an Erd\H{o}s-R\'{e}nyi random graph with average link density $p \in [0,1]$. Without loss of generality, for every $t\in [0,n-2]$, we label the agent $h_0$ with 1 and the agent $l_t$ with $t+2$ in each time period for a parsimonious notation.  In accordance with this labeling rule, there are some agents which are unlabeled in $t<n-1$. In the following analysis, however, we will use fully labeled agents from time period $t=0$ for clear presentation.

\bigskip

\textbf{Network formation in the early stage ($0\leq t< n$).} Initially no link in a network exists, however, individuals perceive that there is a link between any two randomly chosen agents with probability $p$, thus the initial accuracy $\rho_0^i$ for all $i\in N$ is approximated to a normal distribution with mean $1-p+\frac{2}{n}p$. Once the accuracy is revealed in the beginning of time period $t=0$, agent $2$ suggests a link to agent $1$ to improve its network utility by communication with agent 1 for perception update. As seen in (\ref{rho}), the updated perception of agent 2 is more accurate than that of agent $1$ by replacing its perception with $G^1_0$ except the correct information about $n-1$ link states related to agent 2 itself.\footnote{Note that although a link is added between $1$ and $2$ so that $1$ also can communicate with $2$ for obtaining $2$'s own correct information, agent $1$'s perception is not updated because $l_0\neq1$. More generally, all $i\in N\setminus\{l_t\}$ does not change perception details on others. In this paper, we stick on the perception update procedure as specified in (\ref{update}), however, it may be a possible way to expand the model such that connected agents exchange their own correct information.} In $t=1$, agent $3$ adds a link to agent $2$ because $\rho^2_1=\max_{i\in N} \rho^i_1$. Then $3$ is designated as $h$ in $t=2$ so that $4$ offers a link to $3$, and so on. Hence, in the beginning of time period $t<n$, there exist $t$ links, connecting from agent $1$ %and
to $t+2$ in a line network. In time period $t=n-1$, since agent $1$ who has kept its own initial perception becomes the least accurate agent and agent $n$ is the most recently updated agent, $1$ forms a link to $n$ so that a ring structure arises. The early stage dynamics is summarized as %follow:
follows:
\begin{proposition}\label{prop1}
In the early stage ($t< n$), a network evolves in a line in which a branch may be formed with probability $\frac{(1-p)^{n-t-1}}{2}$. In $t=n-1$, a ring structure arises by a link between 1 and n.
\end{proposition}

\begin{figure}[!t]
   \includegraphics[scale=.5]{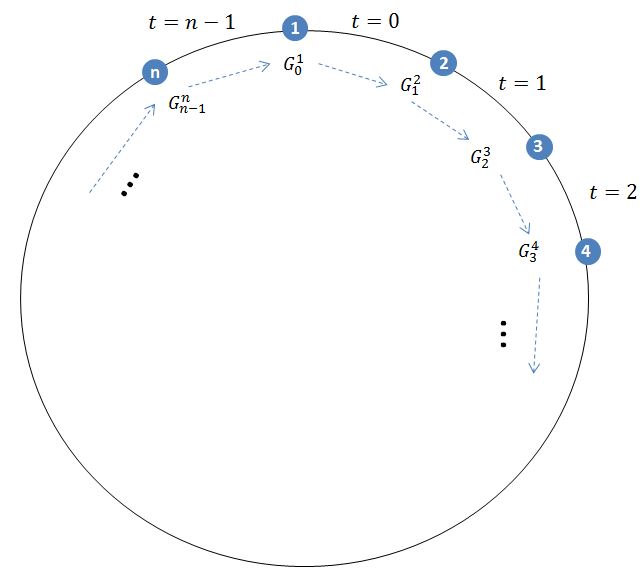} \centering
  \caption{A network in the early stage ($0\leq t<n$)}
    \label{illust}
\end{figure}

Figure \ref{illust} illustrates the formation of a ring network in the early stage. Starting from the link between $1$ and $2$ which is created in $t=0$, a new link is sequentially added to the newly updated agent and a ring structure is completed in $t=n-1$ by a link between $1$ and $n$.\footnote{With a very low probability, it is possible that a network consists of a two-agent component and a growing line component in the early stage $t<n-1$. We do not pay attention on this case because the two-agent component is temporary in the early stage and disappears by connecting with the big component in $t=n-2$. See Appendix in detail.} The dotted arrows denote the flow of information, i.e. the arrow from $G^1_0$ to $G^2_1$ implies that $2$'s perception $G^2_1$ in $t=1$ contains $1$'s original perception $G^1_0$.

A ring structure has a distinguishable property in which connections are not concentrated on any specific agents and the average distance is relatively long.\footnote{Comparing two extreme network structures with the same density, the average distance of a ring network is $\frac{n}{4}$ or $\frac{n}{4}+\frac{1}{2}$, whereas the average distance in a star network (i.e. a single hub node connects all $n-1$ nodes) is less than $2$.} In the simulation section, we verify a ring structure with a few branches by visualizing a network and degree distribution.

\bigskip

\textbf{Perception spreading via links.} When agent $2$ connects with agent $1$ in $t=0$, it replaces own perception with $1$'s perception as described in (\ref{update}). In $t=1$, agent $2$ has $G^2_1$ which contains correct information related to $2$ itself ($e_{2 k,1}$ for all $k \in N$) and $1$'s original perception for other link states ($e^1_{jk,0}$ for all $j, k \in N\setminus\{2\}$) as %follow:
follows:
\begin{eqnarray}
G^1_1 =
 \begin{pmatrix}
  0 &  &  &  \\
  e_{21,1} & 0 &  &  \\
  e_{31,1} & e^1_{32,0}  & 0 \\
  %e_{41,1} & e^1_{42,0}  & e^1_{43,0} \\
  \vdots  & \vdots & \vdots   &  \ddots & \\
  e_{n-1\,1,1} & e^1_{n-1\,2,0}  & e^1_{n-1\,3,0} &\cdots & 0\\
  e_{n1,1} & e^1_{n2,0} & e^1_{n3,0} & \cdots & e^1_{n\,n-1,0} & 0
 \end{pmatrix}, \nonumber \end{eqnarray}
\begin{eqnarray}
 G^2_1 =
 \begin{pmatrix}
  0 &  &  &  \\
  e_{21,1} & 0 &  &  \\
  e_{31,0} & e_{32,1}  & 0 \\
  %e_{41,1} & e_{42,1}  & e^1_{43,0} \\
  \vdots  & \vdots & \vdots   &  \ddots & \\
  e_{n-1\,1,0} & e_{n-1\,2,1}  & e^1_{n-1\,3,0} &\cdots & 0\\
  e_{n1,0} & e_{n2,1} & e^1_{n3,0} & \cdots & e^1_{n\,n-1,0} & 0
 \end{pmatrix}.
 \label{perception}
 \end{eqnarray}
In (\ref{perception}), the only relevant difference between $G^1_1$ and $G^2_1$ is that the second column of $G^1_1$ is $1$'s original perception (i.e. $e^1_{k2,1}=e^1_{k2,0}$ for all $k\in N\setminus\{1\}$), whereas the second column of $G^2_1$ is $2$'s actual link states (i.e. $e^2_{k2,1}=e_{k2,1}$ for all $k\in N$). When agent $2$ updates perception, it mixes $G^1_0$ with own correct information, thus in $G^2_1$, the first column is 1's actual link states which are delivered from 1, the second column is 2's own link states in which 2 knows correctly, and all other columns are 1's original perception which is initially most accurate. Accordingly, $\rho^2_1=\max_{i\in N}\rho^i_1$.\footnote{It is possible that $\rho^2_1=\rho^1_1$ with probability $(1-p)^{n-2}$. More generally, in the early stage $t\in[0, n-1]$, $\rho^{h_t}_{t+1}=\rho^{l_t}_{t+1}$ can happen if $h_t$'s perception on $l_t$'s link states is perfectly correct. Since $h_t$'s perception on $l_t$ comes from $G^1_0$, this event occurs when $e^1_{l_ti,0}=e_{l_ti,t}=0$ for $i=t+3, \cdots, n$. In this case, a branch may be formed because $|H_t|=2$, however perception spreading is unaffected by a branch.} Similarly, once agent $3$ forms a link to agent $2$ in $t=1$, the left three columns in $G^3_2$ are correct information and the rest right $n-3$ columns are $1$'s original perception as shown in (\ref{next_P}). Then $\rho^3_2=\max_{i\in N}\rho^i_2$.
\begin{eqnarray}
 G^3_2 &=&
 \begin{pmatrix}
  0 &  &  &  \\
  e_{21,1} & 0 &  &  \\
  e_{31,2} & e_{32,2}  & 0 \\
  e_{41,0} & e_{42,1}  & e_{43,2} & 0 \\
  \vdots  & \vdots & \vdots  &\vdots &  \ddots & \\
  e_{n-1\,1,0} & e_{n-1\,2,1}  & e_{n-1\,3,2} & e^1_{n-1\,4,0} &\cdots & 0\\
  e_{n1,0} & e_{n2,1} & e_{n3,2} & e^1_{n4,0} & \cdots & e^1_{n\,n-1,0} & 0
 \end{pmatrix}.
 \label{next_P}
 \end{eqnarray}
In general, agents $1, 2,\cdots, t+2$ form a line by sequential linking from $1$ to $t+2$ in the end of time period $t<n-1$. Since each agent's own link states are cumulated when each link is added in a respective time period, agent $t+1$'s perception $G^{t+1}_{t}$ in the beginning of time period $t$ contains the left $t+1$ columns of correct information and the rest right $n-t-1$ columns of $1$'s original perception. Thus, we can derive the highest accuracy in time period $t$ as follows:
\begin{proposition}\label{proposition2}
The highest accuracy in time period $t$ is
\begin{align}
  \rho^{h}_t=\begin{cases}1-\frac{(n-t-1)(n-t-2)}{(n-1)(n-2)}\left(1-\rho^{1}_0\right) &\mbox{ if } t<n-1\\
%\rho^{(1)}_t=\begin{cases}1-\frac{(n-t-1)(n-t-2)}{n(n-1)}\left(1-\rho^{(1)}_0\right) &\mbox{ if } t<n-1\\
1&\mbox{ if } t\geq n-1. \end{cases} \label{highrho}
\end{align}
\end{proposition}

The highest accuracy in each time period $t<n-1$ is a function of the initially highest accuracy $\rho^1_0$ because the initially most accurate perception spreads over connecting individuals in the early stage of dynamics. Once a network is connected in time period $t=n-1$, the initial perception of agent $1$ is exhausted. In a connected network, full information spreads one-by-one with a small adjustment of newly added links.

\bigskip

\textbf{Steady states of a network.} When we started this analysis, we set zero cost to examine an evolving network structure. Now we recall the cost of linking $c>0$ to consider the balance between the benefit of better information and the cost to obtain it. For further analysis, we define a network in a steady state\footnote{We emphasize that the concept of a steady state in this model does not imply cognitive stability. We confine a steady state to network stability that no more new links will be created in a network. Perception update can be continued in a steady state as long as there exists a link between $l$ and $h$.} and the threshold cost from (\ref{case_linear}), as follows:
\begin{definition}\label{def2}
A network reaches a steady state if no agent is willing to create a new link. Formally, a network $G$ is in a steady state if $G_{t+n}=G_{t}$.

In any time period $t$, let $c_t$ denote the threshold cost in which $l$ is not willing to form a link under any cost $c$ higher than or equal to $c_t$, i.e.
\begin{align}
c_t\equiv \frac{2}{n}+\left(1-\frac{2}{n}\right)\rho^{h}_t-\rho^{l}_t. \label{threshold}
\end{align}
\end{definition}
Note that (\ref{threshold}) can be simplified as $c_t=1-\rho^{l}_t$ for $t\geq n-1$ because $\rho^h_t=1$, as shown in Proposition \ref{proposition2}.

Using Definition \ref{def2}, the condition for a steady state is simplified as $c\geq c_t$, and we can characterize network structures in a steady state.

\begin{proposition}\label{proposition3}
If $c\geq c_0$, the network structure in a steady state is an empty network.

Let $\bar{t}$  denote the smallest element in a set $\{\forall t| c_t\leq c_0\}$. If $c< c_0$ and $\bar{t}< n$, there exist $\bar{t}$ links in a steady state. If $c< c_0$ and $\bar{t}\geq n$, there exist at least $n-1$ links in a steady state.
\end{proposition}

Intuitively, Proposition \ref{proposition3} describes that if the cost of linking is sufficiently low for the least accurate agent to initiate a link, the network density in a steady state does not gradually increase, instead, discontinuously jumps from 0 to $\frac{\bar{t}}{M}$ (or higher than $\frac{2}{n}$ if $\bar{t}\geq n$). In particular, consider the early stage dynamics in $t\in [0, n-1]$. The highest accuracy is improved by the amount of individuals' own correct information about $n-t-2$ link states and reaches 1 at the end of the early stage, while the lowest accuracy increases steadily and reaches the initially highest accuracy $\rho^1_0$ at the end of the early stage because the lowest accuracy in each time period is the $(t+1)^{th}$ order statistic of the initial accuracy.%\footnote{The $(t+1)$th order statistic of the initial accuracies and the first order statistic of $n-t-2$ isolated agents' accuracies in time period $t$ may not coincide due to newly added links.}
 Since the amount of individuals' own correct information is larger in the early time steps than in time steps close to $n-1$, the improvement of the highest accuracy is more likely to exceed the improvement of the lowest accuracy which is the difference between $(t+1)^{th}$ and $(t+2)^{th}$ order statistic of $(\rho_0^1, \cdots, \rho_0^n)$ in the early time steps. Accordingly, $c_t$, which is the benefit of obtaining better information, is more likely to increase. Thus, a network in a steady state is connected as long as $c_{n-1}\geq c_0$.

Further evolution depends on the level of $c$ and there is another structural discontinuity that a denser network arises in a steady state if $c<c_{\bar{t}}$. Although there is an analytical limit due to the randomness of linking, we will show it by simulations.

Note that the threshold cost in (\ref{threshold}) indirectly depends on how individuals perceive the initial network. If each individual's perception is not significantly different with each other so that the gap between the highest and lowest accuracy is small, $c_t$ becomes small as well, and a network is less likely to evolve. For instance, given $p$ close to 0, all individuals including the least accurately perceiving agent have almost correct information so that the increment of lowest accuracy by perception update is small, which makes the least accurately perceiving agent hard to initiate a link.

\bigskip

\subsection{Extension}
In the previous subsection, we confined the perception update to a single agent in one time period to discover the early stage of network formation. However, a single agent update may be a strong assumption because a several agents' perception update in one time period can accelerate a structural change. In this subsection, we relax the single agent update assumption as follows:
\begin{assumption}\label{assumption2}
In time period $t$, a chance for the perception update is given to all $i\in L_t$.
\end{assumption}
By allowing a multiple agents' update in one time period, several new links can be formed simultaneously, leading to a different network structure caused by local interactions and a link concentration. Admittedly, precise results about the further evolution after the early stage are not able to be described due to the randomness of linking. Instead, we will provide intuitions why local interactions and a hub agent appear under Assumption \ref{assumption2}. Note that the cost of linking is ignored again (i.e. $c=0$) in order to explore the further evolution of a network.

\bigskip

\textbf{Clustering without a mediator.}  Keeping all other settings the same, we start with the probability that two neighboring agents $k$ and $k+1$ have the same lowest accuracy in  time period $t\in [n, 2n)$.  As discussed in the previous subsection, agent $k$ has less correct information than $k+1$ as $k+1$'s own link states in the early stage, i.e. the $(k+1)^{th}$ column in $G^k_t$ is $(e_{1\,k+1,1},\cdots,e_{k\,k+1,k},0,e^1_{k+2\,k+1,0}, \cdots, e^1_{n\,k+1, 0})$, whereas, the $(k+1)^{th}$ column in $G^{k+1}_t$ is $(e_{1\,k+1,t},\cdots,e_{k\,k+1,t},0,e_{k+2\,k+1,t}, \cdots, e_{n\,k+1,t})$. If agent $1$'s initial perception on $(k+1)$'s link states is correct in time period $t$ ($e^1_{k+2\,k+1,0}=e_{k+2\,k+1, t}=1$ and $e^1_{j\,k+1,0}=e_{j\,k+1,t}=0$ for $j=k+3, \cdots, n$), the two agents have the same accuracy as follows:
\begin{align}
\text{ for } k, k+1 \in N \text{ in time period } t\in [n, 2n), \quad Pr(\rho^k_t=\rho^{k+1}_t)=p(1-p)^{n-k-2}\equiv q_{k}.\label{qkk}
\end{align}

Now we will study the further evolving process after a ring structure has arisen.  An important property of $G_t$ in time period $t\geq n$ is that the newly updated agent neighbors with at least one of the least accurate agents.  At the beginning of time period $t=n$, agent 1's perception $G^1_n$ is most accurate with an update in the previous time period and agent $2$ has the least accurate perception $G^2_n$ which contains $n-2$ columns of $1$'s initial perception as seen in (\ref{perception}). Since the least and the most accurate agents have a link ($e_{12,n}=1$), $2$ updates its perception without the extra cost of creating a new link. Similarly, at the beginning of time period $t=n+1$, agent $3$ who has the lowest accurate perception $G^3_{n+1}$ replaces its own perception with $G^2_{n+1}$ using the link $e_{23,n+1}=1$. This perception update, using an existing link, implies that perception update continues without a change in the network unless $|L_t|>1$. Thus, the existence of multiple least accurate agents is critical in order to form a new link in a ring structure.

Supposing that no new link is created until the two agents $k$ and $k+1$ are in $L_t$ in time period $t=n+k-2$, the full information spreads from agent $1$ to $k-1$ ($\rho^1_t=\cdots =\rho^{k-1}_t=1$). For the perception update, agent $k$ uses an existing link to $k-1$, $e_{k\,k-1,t}=1$, whereas a new link is created between agent $k+1$ and one of agents in $H_t=\{1, \cdots, k-1\}$ with an equal probability. Accordingly, we can derive the probability of a new link as %follow
follows\footnote{Since we are interested in clustering, we ignore the case where the cardinality of $L_t$ is larger than 2.}:
\begin{align}
\prod_{i=1}^{k-1}(1-q_i)q_k.  \label{tie}
\end{align}
Note that when the first new link after the ring structure is created in time period $t=n+k-2$, it is between agent $k+1$ and $h \in H_t=\{1, \cdots, k-1\}$.  %Since the distance between agent $k-1$ and $h$ is limited to $n-(k-1)$, the newly added link between them may not dramatically reduce the average distance of a network, unlike the well-known random rewiring model by Watts and Strogatz (1998).
Especially, the more frequently new links are formed, the smaller $H_t$ is, implying that new links are more likely to be a local interaction rather than a short cut.\footnote{Note that the probability (\ref{tie}) is reduces by half under the single agent update assumption because a new link is formed only if $k+1$ is selected as $l$. Since a new link is rarely added relative to the multiple agents' perception update case, $H_t$ is larger and the new link is more likely to be a short cut in the single agent update case.}

It is worth emphasizing the idea in the previous paragraph that clusters in a social network can be formed without a mediator. Studies by V\'{a}zquez (2003) and Jackson and Rogers (2007) explain triadic relations in which one connects with a new node through existing connections, thus any two nodes sharing a common neighbor are more likely to be connected. Our result adds another clustering mechanism which does not need a mediating node: if information transmission continues through a link, a cluster is formed because the most/least well perceiving agents are closely located. If individual agents' perception accuracy is similar so that minimum accuracy ties frequently appear, local interactions actively occur, leading to more clusters. Since local interactions and clusters do not dramatically reduce the average distance, we shall show \emph{not-so-small-world} in the perceptual attachment\footnote{We name this link formation process as ``\emph{perceptual attachment}''. The name ``perceptual attachment'' emphasizes that new links are added based on subjective cognition rather than correct information, unlike Barab\'asi-Albert model (i.e. preferential attachment) assumes.} process with simulations.

\bigskip

\textbf{Link concentration.}  As we have analyzed in the previous subsection, agent 1 and $n$ have full information in time period $t=n-1$ and the full information with small fluctuations caused by new links spreads throughout individuals in $t\in[n,2n)$. After $t=2n$, agents have almost correct information and their accuracies vary only within a thin range close to 1. Now we discuss a link concentration in the further evolution. 

\smallskip

In time period $t>2n$, most agents have at least two links in a global\footnote{We describe a network with the word ``global'' because local connections do not change the overall ring structure so that the ring structure is sustained in the further evolving process.} ring network with local connections. While individuals share correct information about the ring structure which is completed at the end of the early stage, the information about new links which are formed in $t\in[n, 2n]$ is known to related agents only. Due to the common perception on the ring structure, accuracy ties more frequently occur in $t>2n$ and by Assumption \ref{assumption2}, all agents in $L_t$ are allowed to update their perception.

Suppose that $|L_t|>2$ and all agents in $L_t$ except agent $i$ and $j$ have a link to any $h\in H_t$ (i.e. for all $l\in L_t\setminus\{i, j\}$, $e_{lh,t}=1$ for any $h\in H_t$ and $e_{ih,t}=e_{jh,t}=0$ for $i, j\in L_t$ and all $h\in H_t$). For a perception update, $i$ and $j$ need to form a link to any agents in $H_t$. Let $h^i$ and $h^j$ denote the agent in $H_t$ chosen to be linked by $i$ and $j$ respectively. As shown in (\ref{rhoUpdate}), the links between $i$ and $h^i$ and between $j$ and $h^j$ in time period $t$ reduce accuracies of all agents except the related agents in the next time period because
\begin{align}
&e^k_{ih^i,t+1}=e^k_{ih^i,t}=e_{ih^i,t}=0 \neq e_{ih^i,t+1}=1 \text{ for all }k\in N\setminus\{i,h^i\}, \nonumber\\
&e^k_{jh^j,t+1}=e^k_{jh^j,t}=e_{jh^j,t}=0 \neq e_{jh^j,t+1}=1 \text{ for all }k\in N\setminus\{j,h^j\}. \nonumber
\end{align}
For a link concentration, we consider the case where $i$ and $j$ accidentally choose the same agent in $t$, i.e. $h^i=h^j\equiv h^*$. By (\ref{update}) for an update, $i$ and $j$ have the information inherited from $h^*$ and their own link states which are unknown to each other in the next time period. Thus, although $h^*$, $i$, and $j$ have more accurate information than all other agents in time period $t+1$, $i$'s perception on $j$ and $h^*$'s link becomes wrong due to the new link between $j$ and $h^*$ in $t$ (i.e. $e^i_{jh^*,t+1}=e^{h^*}_{jh^*,t}=e_{jh^*,t}=0 \neq e_{jh^*,t+1}=1$), and $j$ also misperceives the link between $i$ and $h^*$. On the other hand, since $h^*$ is involved in the two new links, $h^*$'s accuracy is strictly higher than $i$ and $j$, leading to $h^*$ as a single element in $H_{t+1}$.

Although $h^*$ becomes the most accurate agent by chance, all new links after this time period must be involved with $h^*$: In time period $t+1$, all $i\in L_{t+1}$ update their perception with $G^{h^*}_{t+1}$. If new links are added for the update, the new links ensure $h^*$'s unique highest accuracy in the next time period $t+2$ as the information about the new links is partially known to those who offer the links to $h^*$, whereas $h^*$ fully observes the new links.

It is worth mentioning a link concentration mechanism in the preferential attachment process (Barab\'{a}si and Albert, 1999). The preferential attachment results in the same conclusion that an agent who obtains more links by chance is highly likely to grow into a hub in a growing network. However, the mechanism is fundamentally different in regards to whether the linking process is stochastic or strategic: In the preferential attachment process, obtaining a new link is the matter of probability such that the more connected agent has the higher probability to get a new link, whereas, in our model, the more connected agent is meant to be a hub due to the exclusive information about its own links.

\bigskip

\section{Simulation}
\label{simulation}

In this section, we perform numerical simulations of the model by setting $n=200$ and $p=0.1$, while the cost of linking $c$ is the only relevant parameter. Initially, the network is empty and the perception accuracy of the network is normally distributed around $E[\rho_0]=\frac{2}{n}+(1-\frac{2}{n})(1-p)=0.901$.

In the previous section, we discussed how a network evolves in the early stage, how perception spreads, when a network reaches a steady state, and how the frequency of linking affects network structures. In this section, we examine these %hypothesis
hypotheses one by one.

\bigskip

\begin{hypothesis} The network evolves in a line and a ring structure arises at the end of the early stage. \end{hypothesis}
\begin{figure}[h]
\includegraphics[width=\columnwidth]{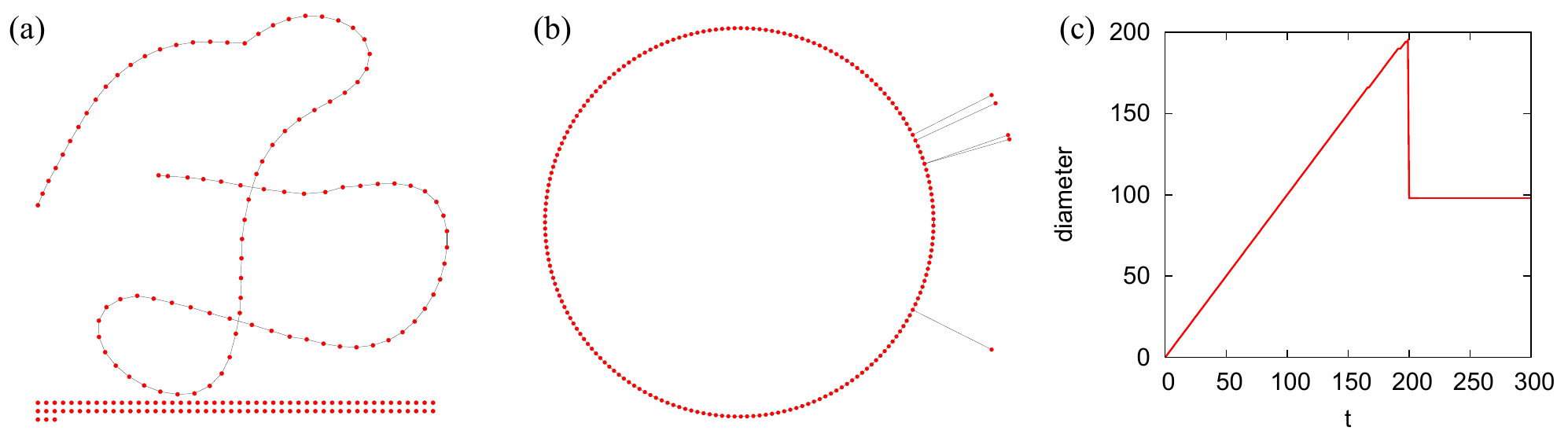}\centering
    \caption{Networks (a) at $t=100$ and (b) at $t=199$. (c) The network diameter.}
    \label{hypo1}
\end{figure}

Firstly, network snapshots in Figure \ref{hypo1} (a) and (b) visualize how a network evolves in the early stage. Figure \ref{hypo1} (a) shows a line network before being connected in $t=100$ and (b) shows a ring structure when it is connected in $t=199$. Moreover, the network diameter\footnote{By definition, the diameter of a network is the longest path length.} in (c) also reveals the growing line structure for $t<200$: the diameter linearly increases with time period $t$, implying that the network is expanding by a new link to the end agent in each time period. Observe that in time period $t=199$, the diameter is reduced by half as the ring structure is completed.

\bigskip

\begin{hypothesis} Accumulation of each individual's own correct information increases the highest accuracy in the early stage and full information spreads after $t=n-1$.\end{hypothesis}

\begin{figure}[h]
\includegraphics[width=.6\columnwidth]{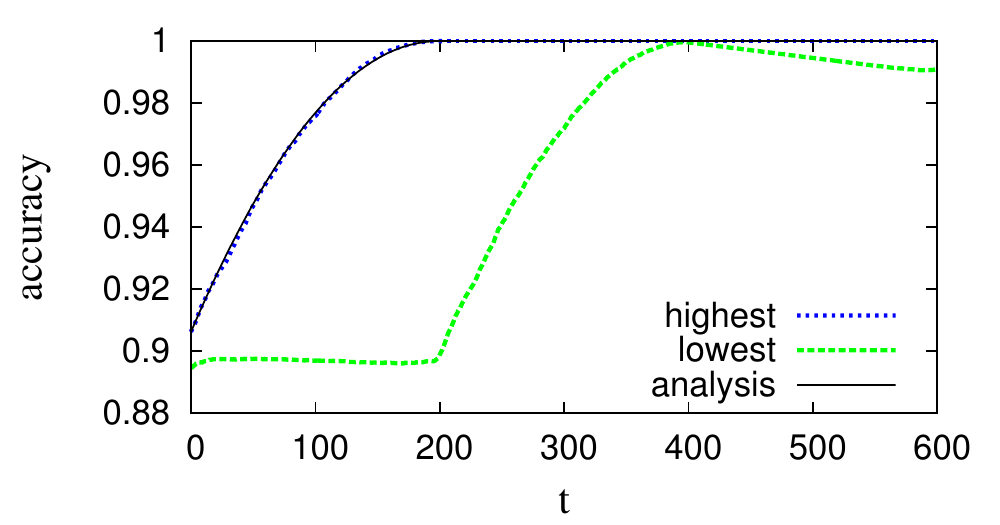}\centering
    \caption{Evolution of perception accuracies: $\rho^h_t$ and $\rho^l_t$.}
    \label{hypo2}
\end{figure}
In Proposition \ref{proposition2}, we derived the highest accuracy in %each
time period $t$. The highest accuracy in the early stage $t\in[0,n-1]$ depends on the initially highest accuracy until the initially most accurate perception has been completely replaced with correct information. To verify this result, Figure \ref{hypo2} illustrates the highest and lowest accuracy as a function of time. Given $\rho^h_0=0.906$, the black curve shows $\rho^h_t$ for all $t$ in (\ref{highrho}), accompanying with corresponding numerical results denoted by the blue dotted curve. The green dotted curve denotes $\rho^l_t$. For the lowest accuracy, $\rho^l_t$ in the early stage roughly reflects the initial distribution of $\rho^i$ in descending order. In the second round $t\in [n, 2n)$, $\rho^l_t$ simply corresponds to $\rho^h_{t-n}$ due to the sequential update of perception.

\bigskip

\begin{hypothesis} There is a discontinuous jump of the network density in a steady state.\end{hypothesis}

Since $\rho^h_0=0.906$ and $\rho^l_0=0.895$, from (\ref{threshold}), the condition for non-empty network is $c<c_0=0.01230$. When $c\geq c_0$, the network in a steady state is an empty network because even the very first link will not be added. If the first link is formed for the cost $c=c_0-\varepsilon$, where $\varepsilon$ is a positive and arbitrarily small number, the network reaches a steady state in $t=333$ when $c_0=c_{333}$. It indicates a sharp transition from zero link density ($c\geq c_0$) to a finite density ($c<c_0$) in a steady state.

\begin{figure}[h]
\includegraphics[width=\columnwidth]{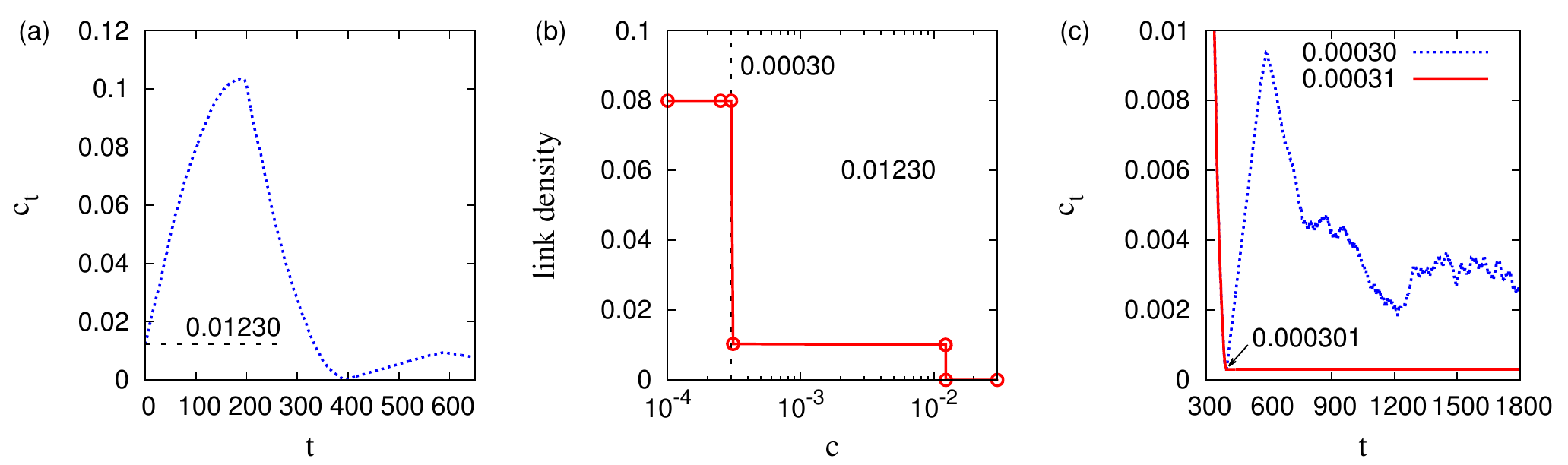}\centering
    \caption{(a) Threshold cost $c_t$. (b) Network density and the cost of linking. (c) $c_t$ in detail for $0.00030$, and $0.00031$.}
    \label{hypo3}
\end{figure}

To examine the density transition in detail, Figure \ref{hypo3} illustrates the threshold cost and the network density for given values of $c$. In (a), the threshold cost $c_t$ is increasing in the early stage, which indicates that if the cost of linking $c$ is low enough to initiate the first link, the network in a steady state is connected. In particular, since $\bar{t}=333$, once the first link is added, the network continues evolving by $t=333$ so that there exist at least 199 links in a steady state. The graph in (b) is the network density in accordance with the cost of linking. We verify the structural discontinuity in (b) that the link density is zero for $c>0.01230$. This graph shows another sharp transition of the link density at $c=0.00030$. This jump comes from fluctuations in $c_t<c_0$. To investigate the fluctuations, the graph in (c) compares $c_t$ at this transition point. The blue and red curves denote $c_t$ given the cost $c=0.00030$ and $c=0.00031$ respectively. At the cost $c=0.00031$, $c_t$ is horizontal flat after $t=397$, which implies that the network stops evolving. On the contrary, at the slightly lower level of cost $c=0.00030$, the network continues evolving and the threshold cost $c_t$ shows an increasing trend from $t=397$. With the same reason for the sharp transition at $c=c_0$, we can expect another sharp transition of the network density at this level of cost: Letting $\hat{c}$ be $c_{397}=0.000301$, the network stops evolving if $c\geq \hat{c}$ or continues evolving if $c<\hat{c}$. Hence, if $c=0.00031>\hat{c}$, no links are added after time period $t=397$, while perception spreads along existing links. On the other hand, if $c=0.00030<\hat{c}$, the network continues evolving.

\begin{figure}[!h]
\includegraphics[width=\columnwidth]{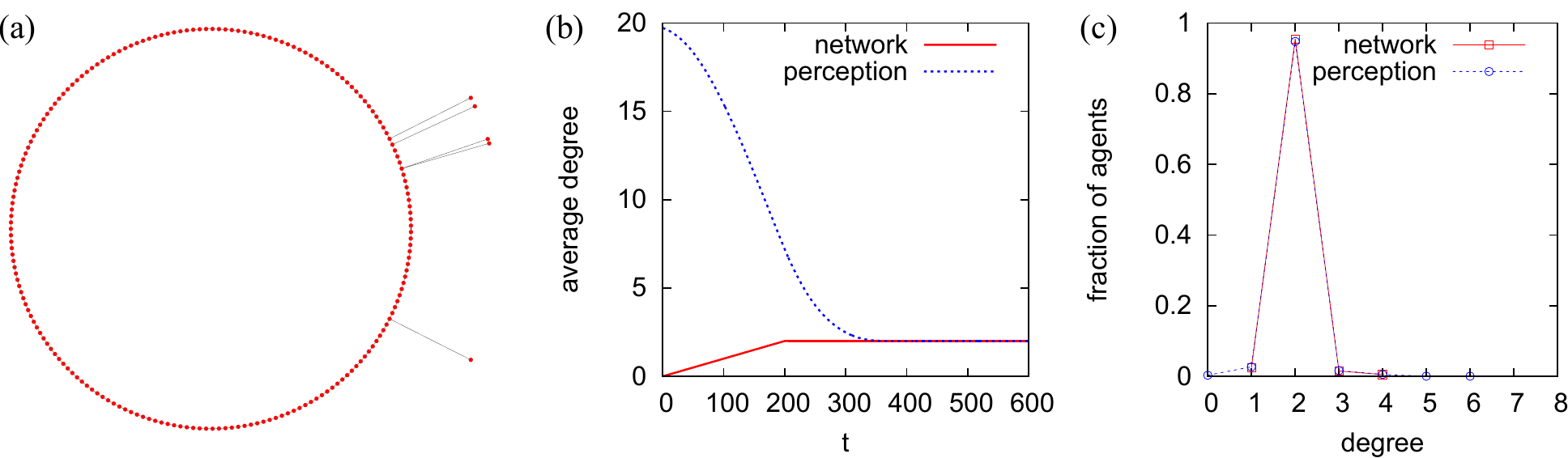}\centering
    \caption{For $c=0.0122$, (a) the network in the steady state, (b) the evolution of average degree in the network and in perceptions, and (c) degree distributions of the network and perceived networks in a steady state.}
    \label{ssntw1}
\end{figure}

\begin{figure}[!h]
\includegraphics[width=.666\columnwidth]{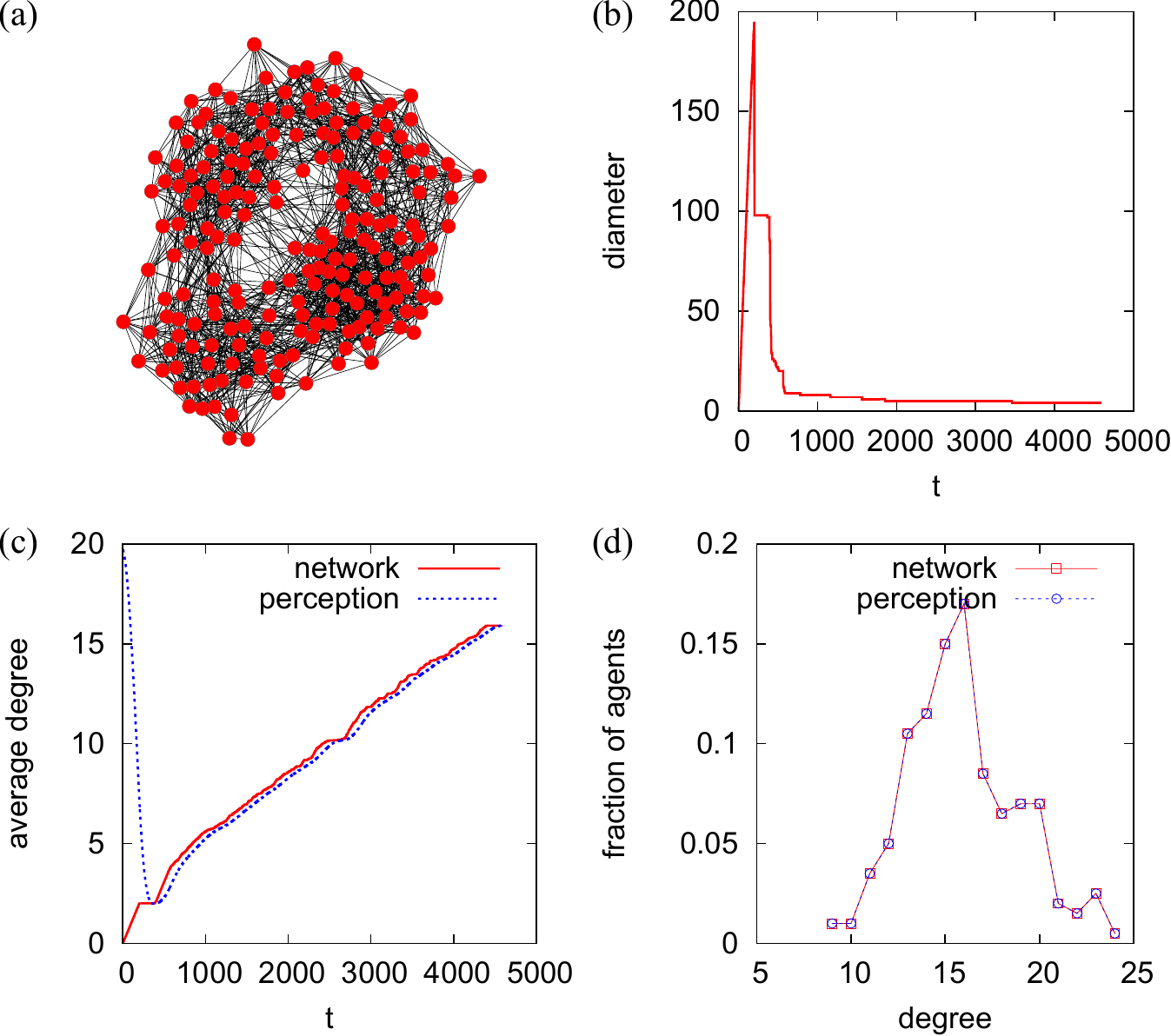}\centering
    \caption{For $c=0.00030$, (a) the network in the steady state, the evolution of (b) diameter and (c) average degrees in the network and in perceptions, and (d) degree distributions of the network and perceived networks in a steady state.}
    \label{ssntw2}
\end{figure}

For a further comparison of network structures at the different level of cost, we illustrate two cases in Figures \ref{ssntw1} and \ref{ssntw2}. Firstly, Figure \ref{ssntw1} (a) depicts the network in a steady state for the high cost of linking, i.e. $c=0.0122\in[\hat{c}, c_0]$. The overall structure is a ring with a few branches. Figure \ref{ssntw1} (b) is the evolution of the average degree, showing how perception coevolves with the network structure. The average degree of individual agents' perception, denoted by the blue dotted curve, quickly converges to the average degree of the actual network. Note that although no new links are formed after $t=199$, perception updates continue via existing links until all agents have almost correct information, which is shown in (c) as both degree distributions of perceptions and the actual network are close to each other.

Secondly, Figure \ref{ssntw2} shows the network in a steady state for the low cost of linking, i.e. $c=0.00030 <\hat{c}$. The low cost of linking enables a continuous change in the network structure by adding links, thus the full information under the low cost is temporary. If a network continues evolving due to the low cost of linking, individuals are more willing to update perception because their information becomes incorrect quickly, leading to a higher benefit of link creation. Thus, the least accurate agent more easily adds a link to the most accurate agent, which makes others' perception more inaccurate. Figure \ref{ssntw2} (c) shows how perception converges to the actual network, with a little gap between perception and the actual network which encourages agents to add more links, as explained.

In time periods $t\in[n, 2n]$, since the probability $q_{k}$ in (\ref{qkk}) is low for low labeled agents, the correct information about the network spreads over many agents. Whereas, $q_k$ is relatively high for high labeled agent, new links are formed as $t$ closes to $t=2n$, and these new links are more likely to be a short cut due to the correct information spreading. Figure \ref{ssntw2} (b) verifies short cuts by showing the dramatically decreasing diameter. These short cuts significantly change the ring structure, and the structural change is accelerated in the further evolving process in $t>2n$. Eventually, a very dense complicated structure in Figure \ref{ssntw2} (a) arises in a steady state and degree distribution in Figure \ref{ssntw2} (d) shows a relatively symmetric degree distribution.

\bigskip

\begin{hypothesis} If all $l\in L_t$ are allowed to update perception, new links are more likely to be a local interaction.\end{hypothesis}

\begin{figure}[h]
\includegraphics[width=.666\columnwidth]{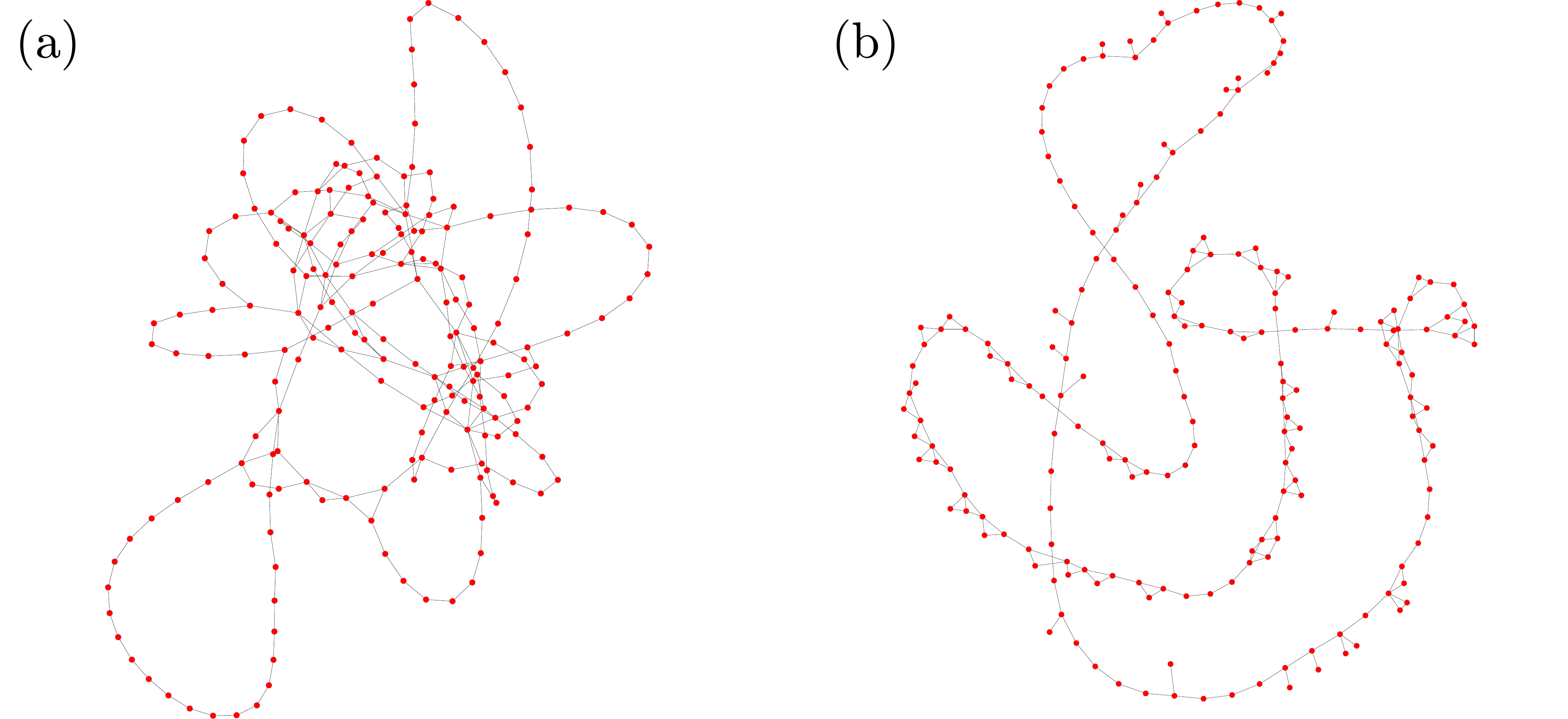}\centering
    \caption{Networks with 250 links  (a) for the single agent's update and (b) for multiple agents' update.}
    \label{hypo4}
\end{figure}

As discussed in Extension, if the single agent's update assumption is relaxed to allow multiple links to be created in one time period, we expect that the probability of creating short cuts will be very low compared to the single agent's update case. Thus once the network is connected, it keeps the global ring structure with local connections. To see how the frequency of perception updates in one time period affects a network structure, we compare two networks which have the same density (250 in each network) but one is formed by a single agent's update in Figure \ref{hypo4} (a) and the other is formed by a multiple agents' update in Figure \ref{hypo4} (b). The two networks are structurally different: In (a), there are many short cuts which destroys the ring structure. In (b), there are many triangle connections in the ring network. We consistently find that the clustering coefficient of the network in (b) is $0.287$, which is higher than $0.024$ for the network in (a).

Intuitively, we can interpret this result that people in a rapidly changing network are more likely to interact locally because the newest information is more accurate, whereas in a slowly changing network, information spreads throughout plenty of people so that interactions are not necessarily local.

\bigskip

\begin{hypothesis} If all $l\in L_t$ are allowed to update perception, a slightly more connected agent grows into a hub.\end{hypothesis}

\begin{figure}[h]
\includegraphics[width=.666\columnwidth]{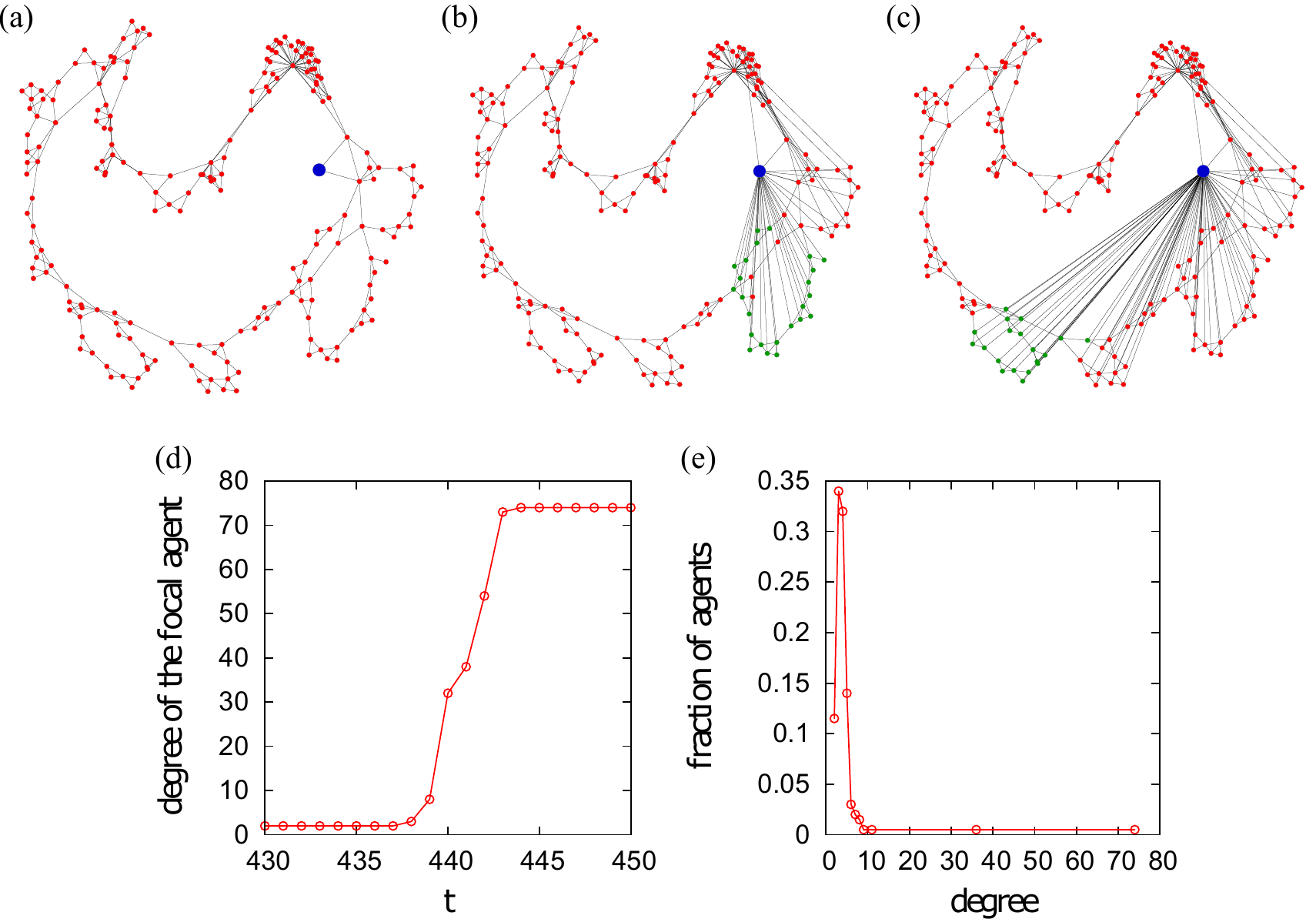}\centering
\caption{Network snapshots for the focal agent (a) in $t=437$, (b) $t=440$, and (c) $t=443$. (d) Degree trend of the focal agent and (e) degree distribution of the network in $t=444$.}
    \label{hypo5}
\end{figure}

Figure \ref{hypo5} shows the emergence of a hub agent denoted by a big blue node in a network. We trace this agent's degree trend over time. In (a), like other agents, the degree of this agent remains $2$ until $t=437$. Once this agent updates its accuracy and becomes the most accurate agent in $t=438$, it attracts $6$ out of $13$ lowest accurate agents by chance. After the degree increases to $8$, this agent becomes the single most accurate agent (i.e. $|H_t|=1$) in $t=439$ to $443$ so that all the least accurate agents who is unconnected with this node form a link, emerging as a hub with degree $74$. The two network snapshots in (b) and (c) illustrate this agent and all agents in $L_t$ denoted by green nodes. The degree trend in (d) verifies this process: in $t=438$, the degree of this agent starts surging and reaches 74 just within a few time steps. The degree distribution of the network in (e) reveals a high link concentration.

\bigskip

\section{Conclusion}
\label{conclusion}
This paper has proposed a simple model to study the coevolution between a network and perception. Focusing on how individuals and a system affect each other, we have examined a cognition-based strategic link formation. Assuming that a link as a conduit of communication is costly and more accurate perception yields higher network utility, one decides whether to form a link in order to get better knowledge. A newly added link causes a change in a network, which affects an individual's perception accuracy back.

We found that a network evolves in a line in the early stage and a ring structure arises once connected. Due to correct information added by each individual, the highest accuracy is improved and there must be an agent who possesses the full information in a connected network. We also showed discontinuous network density in a steady state and observed local interactions and a link concentration in a frequently changing network, which provides a plausible reasoning for clusters and a hub in a social network. Additionally, the relationship between an evolving process and the cost of linking has been discussed and a simulation illustrated how a network and perception coevolve.

This research is meaningful by revealing the importance of cognition in the coevolution between individuals and a system. As we have suggested a simplified framework of the interplay between an individual's perspective and a systemic change, there are potential ways to develop further models in psychology, economics, and sociology. We retain extensions of this model to various directions including an experiment for the future research.

\bigskip

%\singlespacing

\appendix
\section*{Appendix}
\normalsize
\textbf{Proposition \ref{prop1} }\begin{proof}
  Suppose that $|H_t|=1$ for all $t\in[0, n-1]$. In $t=0$, the unique element of $H_0$ is $h_0=1$, and $l_0$, labeled with $2$, forms a link to $h_0$, i.e. $e_{12,0}=1$. For $t>0$, by (\ref{rho}), $\rho^{l_{t-1}}_t=\max_{i\in N}\rho^i_t$ so that the unique element of $H_t$ is $h_t=l_{t-1}=t+1$. In time period $t$, $l_t$, labeled with $t+2$, forms a link to $h_t$, labeled with $t+1$, i.e. $e_{t+1\,t+2,t}=1$. In $t+1$, by (\ref{rho}), $\rho^{l_t}_{t+1}=\max_{i\in N}\rho^i_{t+1}$ so that $h_{t+1}=l_t=t+2$ and a link is formed between $t+1$ and $l_{t+1}$, labeled with $t+3$, thus $$e_{t+1\,t+2,t}=e_{t+2\,t+3,t+1}=1.$$ Growing in a line, this process continues until $t=n-1$ when $h_{n-1}=n$ and $l_{n-1}=1$ which completes a ring structure by connecting both end of the line.

\smallskip

Suppose that $|H_t|\geq 2$. The randomly chosen element in $H_0$ will be connected by $l_0$, labeled with $2$. For $i\in H_0\setminus\{h_0\}$, if $\rho^{i}_1<\max (\rho^{l_0}_1, \rho^{h_0}_1)$, a new link in $t=1$ is added to the more accurate agent between $l_0$ and $h_0$, in which a network continues evolving in a line.  If $\rho^{i}_1=\rho^{h_0}_1=\rho^{l_0}_1$ by an accidental correction of $i$'s accuracy and either $l_0$ or $h_0$ is chosen as $h_1$, a network is expanded in a line. If $\rho^{i}_1=\rho^{h_0}_1=\rho^{l_0}_1$ and $i=h_1$, $i$ and $l_1$ are linked in $t=1$, which are separated from the component of agent $h_0$ and $l_0$. When perception of $n-2$ agents from $l_1$ to $n$ has been updated and the perception update chance is given to $l_0$ again in $t=n-2$, the network is connected by the link between $l_0$ and $n$.

Then we consider the case with $|H_t|\geq 2$ for $t>0$. Firstly, for $|H_t|=2$, $H_t=\{l_{t-1}, h_{t-1}\}=\{t+1, t\}$ by $e^t_{i\,t+1,t}=e^1_{i\,t+1,0}=e_{i\,t+1,t}=0$ for all $i\in[t+2,n]$. In this case, a branch is formed only if $h_t=h_{t-1}=t$, thus, the probability that a branch is created in $t$ is derived as $$\Pr(h_t=h_{t-1})\prod_{i=t+2}^{n}\Pr\left(e^1_{i\,t+1,0}=0\right)=\frac{(1-p)^{n-t-1}}{2}.$$
Now the proof is completed by showing that $|H_t|>2$ does not happen for $t>1$. %Since $|H_{t+1}|>1$ in time $t+1$, $\rho^{l_t}_{t+1}=\rho^{h_t}_{t+1}$, equivalently to $\rho^{n-t}_{t+1}=\rho^{n-t+1}_{t+1}$.
If $|H_t|>2$, it must be the case that $|H_{t-1}|>1$ and $|H_t|>1$ so that
\begin{align}
  \rho^{l_{t-2}}_{t-1}=\rho^{h_{t-2}}_{t-1}\text{ and }\rho^{l_{t-1}}_{t}=\rho^{h_{t-1}}_{t} \quad\Leftrightarrow\quad & \rho^{t}_{t-1}=\rho^{t-1}_{t-1} \text{ and } \rho^{t+1}_{t}=\rho^{t}_{t} \nonumber
\end{align}
Since $|H_t|>2$,
\begin{align}
\rho^{t-1}_{t}=\rho^{t}_{t}=\rho^{t+1}_{t}.\label{twoMax}
\end{align}
For $\rho^{t-1}_{t}=\rho^{t}_{t}$, $e^{t}_{t-1\,t+1,t-1}=e^{t-1}_{t\,t+1,t-1}=0$ because
\begin{align}
&e^{t}_{t-1\,t+1,t-1}=e^{t-1}_{t-1\,t+1,t-1}=e_{t-1\,t+1,t-1}=0\text{ and }\nonumber\\
&e^{t-1}_{t\,t+1,t-1}=e^{t}_{t\,t+1,t-1}=e_{t\,t+1,t-1}=0.\nonumber
\end{align}
In time period $t-1$, if $h_{t-1}=t$, $e^{t-1}_{t\,t+1,t-1}\neq e^{t}_{t\,t+1,t-1}=e_{t\,t+1,t-1}=1$ so that $\rho^{t}_{t}>\rho^{t-1}_{t}$, which contradicts (\ref{twoMax}). Similarly, if $h_{t-1}=t-1$, $e^{t}_{t-1\,t+1,t-1}\neq e^{t-1}_{t-1\,t+1,t-1}=e_{t-1\,t+1,t-1}=1$ so that $\rho^{t-1}_{t}>\rho^{t}_{t}$, which also contradicts (\ref{twoMax}).
\end{proof}

\bigskip

\textbf{Proposition \ref{proposition2} }\begin{proof}
  In time period $t=0$, by definition, $\rho^{h_0}_0=\rho^1_0$. In $t>0$, the perception of $h_t=t+1$ contains the left $t$ columns of correct information which have been added up by agents $1, \cdots, t$, and the $(t+1)^{th}$ column of correct information by $t+1$ itself so that $t+1$'s perception becomes most accurate with the left $t+1$ columns of correct information and the rest right $n-t-1$ columns of initial perception of $h_0$. Accordingly, $\rho^{h_t}_t$ can be decomposed with correct information and $e^1_{jk,0}$ as follows:
\begin{align}
  G^{h_t}_t=G^{t+1}_t=\begin{pmatrix}
  0 &  &  &  & & &\\
  e_{21,1} & 0 &  &  \\
  e_{31,2} & e_{32,2}  & 0\\
  \vdots  & \vdots   &  \vdots \\
  e_{t+1\,1,t} & e_{t+1\,2,t}  & \cdots & 0\\
  e_{t+2\,1,0} & e_{t+2\,2,1}  & \cdots & e_{t+2\,t+1,t} & 0\\
  e_{t+3\,1,0} & e_{t+3\,2,1}  & \cdots & e_{t+3\,t+1,t} & e^1_{t+3\,t+2,0} &0\\
  \vdots & \vdots & & \vdots&\\
  e_{n1,0} & e_{n2,1} & \cdots & e_{n\,t+1,t} & e^1_{n\,t+2,0} & \cdots & e^1_{n\,n-1,0} & 0
 \end{pmatrix} \nonumber
\end{align}
\begin{align}
  \rho^{h_t}_t=\frac{1}{M}\left\{r^1\sum_{k=1}^{n-t-2}k + 1\left(M-\sum_{k=1}^{n-t-2}k \right)\right\}=1-\frac{(n-t-1)(n-t-2)}{(n-1)(n-2)}\left(1-\rho^1_0\right). \label{leftover}
\end{align}
Here let $r^i$ denote the average of $I(e^i_{jk,0},0)$ for all $j,k\in N\setminus \{i\}$:
\begin{align}
r^i\equiv \frac{1}{M-n+1}\sum_{j\in N\setminus\{i\}}\sum_{k\in N\setminus\{i\}, k>j}I(e^{i}_{jk,0},0)=\frac{\rho_0^i-2/n}{1-2/n}.\label{ri}
\end{align}
Equation (\ref{leftover}) satisfies $\rho^{h_0}_0=\rho^1_0$. If $t=n-1$, $\rho^{h_{n-1}}_{n-1}=1$ in (\ref{leftover}). Since $l$ obtains the full information from $h$ after $t=n-1$, $\rho^{h_t}_t=1$ for all $t\geq n-1$.
\end{proof}

\bigskip

\textbf{Proposition \ref{proposition3}} \begin{proof}
Firstly, if $c\geq c_0$, no link will be initiated so that the network is empty.

\smallskip

Suppose $c=c_0-\varepsilon$ for a sufficiently small positive $\varepsilon$ so that $c$ is close to the threshold cost $c_0$ but a link between $l_0$ and $h_0$ is created. In $t=1$, if $c\approx c_0<c_1$, $l_1$ forms a link to $h_1$ and a network continues evolving in the next time period. If $c\geq c_1$, no link is added after $t=1$ because there is no change in a network and the condition for a new link cannot be satisfied without a new change, thus $\bar{t}=1$ and one link exists in this network. Similarly, repeating the same argument by $t-1$, suppose that $c_t<c\approx c_0\leq c_{t-1}$ and there are $t$ links which have been added in each period from 0 to $t-1$. In time period $t$, since $c>c_t$, $l_t$ does not form a link and no change occurs in a network, which indicates that the network is in a steady state.

\smallskip

In $t\in[0, n-1]$, $c<c_t$ is a necessary and sufficient condition for a new link because $e_{lh,t}=0$ in the early stage. Thus, if $\bar{t}\leq n-1$, there exist $\bar{t}$ links in a steady state. In $t>n-1$, $c<c_t$ is a necessary condition for a new link because $l$ and $h$ can be already linked. Thus, if $\bar{t}>n-1$, there exist at least $n-1$ links in a steady state.
\end{proof}

\end{document}